\documentstyle[11pt,twoside]{article}
\def\greaterthansquiggle{\raise.3ex\hbox{$>$\kern-.75em\lower1ex\hbox{$\sim$}}}
\def\lessthansquiggle{\raise.3ex\hbox{$<$\kern-.75em\lower1ex\hbox{$\sim$}}}
\newcommand{\beq}{\begin{equation}}
\newcommand{\eeq}{\end{equation}}
\newcommand{\beqa}{\begin{eqnarray}}
\newcommand{\eeqa}{\end{eqnarray}}
\newcommand{\beqan}{\begin{eqnarray*}}
\newcommand{\eeqan}{\end{eqnarray*}}
\newcommand{\ba}{\begin{array}}
\newcommand{\ea}{\end{array}}
\newcommand{\no}{\nonumber}

\newcommand{\ra}{\rightarrow}

\newcommand{\ve}{\varepsilon}

\newcommand{\dg}{\dagger}
\newcommand{\wt}{\widetilde}

\newcommand{\N}{{\cal N}}

\def\nz{\ifmmode {I\hskip -3pt N} \else {\hbox {$I\hskip -3pt N$}}\fi}
\def\zz{\ifmmode {Z\hskip -4.8pt Z} \else
       {\hbox {$Z\hskip -4.8pt Z$}}\fi}
\def\qz{\ifmmode {Q\hskip -5.0pt\vrule height6.0pt depth 0pt
       \hskip 6pt} \else {\hbox
       {$Q\hskip -5.0pt\vrule height6.0pt depth 0pt\hskip 6pt$}}\fi}
\def\rz{\ifmmode {I\hskip -3pt R} \else {\hbox {$I\hskip -3pt R$}}\fi}
\def\cz{\ifmmode {C\hskip -4.8pt\vrule height5.8pt\hskip 6.3pt} \else
       {\hbox {$C\hskip -4.8pt\vrule height5.8pt\hskip 6.3pt$}}\fi}

\def\au{{\setbox0=\hbox{\lower1.36775ex%
\hbox{''}\kern-.05em}\dp0=.36775ex\hskip0pt\box0}}
\def\ao{{}\kern-.10em\hbox{``}}

\normalsize
\begin{document}
\bibliographystyle{plain}
\begin{titlepage}
\begin{flushright}
UWThPh-1994-40\\
\today
\end{flushright}
\vspace{2cm}
\begin{center}
{\Large \bf
Transition Radiation of Ultrarelativistic Neutral Particles*}\\[50pt]
W. Grimus and H. Neufeld  \\
Institut f\"ur Theoretische Physik \\
Universit\"at Wien \\
Boltzmanngasse 5, A-1090 Wien
\vfill
{\bf Abstract} \\
\end{center}

We perform a quantum theoretical calculation of transition radiation by
neutral particles with spin 1/2 equipped with magnetic moments and/or
electric dipole moments. The limit of vanishing masses is treated exactly
for arbitrary refraction index. Finally we apply our result to the solar
neutrino flux.
\vfill
\noindent * Supported in part by Jubil\"aumsfonds der
\"Osterreichischen Nationalbank, Project~No.~5051
\end{titlepage}

Neutrinos seem to be likely candidates for carrying features of physics
beyond the Standard Model. Apart from masses and mixings also magnetic
moments (MM) and electric dipole moments (EDM) are signs of new physics
and are of relevance in terrestrial experiments, the solar neutrino
problem \cite{ci}, astrophysics and cosmology. Elastic $\nu$--$e^-$
scattering restricts MMs to $\mu^2_{\nu_e} + 2.1 \mu^2_{\nu_\mu} <
1.16 \cdot 10^{-18} \mu^2_B$ \cite{kr} and
$|\mu_{\nu_\tau}| < 5.4 \cdot 10^{-7} \mu_B$ \cite{co} where
$\mu_B$ is the Bohr magneton. Limits on $\mu_{\nu_\tau}$ from
$e^+ e^- \ra \nu \bar \nu \gamma$ are an order of magnitude weaker
\cite {gro}. Note that in the above inequalities $\mu^2$ can be replaced
by $\mu^2 + d^2$ since MMs $\mu$ and EDMs $d$ become indistinguishable
when neutrino masses can be neglected. Astrophysical and cosmological
bounds are considerably tighter but subject to additional assumptions
(see e.g. \cite{ra}).

In this paper we explore the possibility of using electromagnetic
interactions of neutrinos with polarizable media to get information on
neutrino MMs and EDMs. The case of an infinitely extended homogeneous
medium where Cherenkov radiation is emitted has already been discussed
in a previous paper \cite{gri}. Here we concentrate so to speak on the
opposite situation, namely a sharp boundary between two different
homogeneous media where transition radiation \cite{gif} is emitted when
a particle crosses such a boundary.
In contrast to Cherenkov radiation the particle can radiate at any
non--zero velocity in this case. Classical discussions of transition
radiation are e.g. found in \cite{git,du} whereas the quantum theoretical
approach has been used in \cite{ga,be}. In this paper we will perform
a calculation for general neutral spin 1/2 particles with MMs and EDMs.
As pointed out in \cite{ga}, the quantum theoretical treatment of
transition radiation of the electron changes the classical result very
little but, on the other hand, is essential in our case as will become
clear later. Our aim is not only to estimate the transition radiation
caused by the solar neutrino flux but also to exhibit the general
methods of our calculation and successive approximations. We think
therefore that the present work can also be used as a basis for
considering transition radiation in the case of e.g. non--relativistic
particles, large diffractive indices etc. where the usual computations
are not applicable.

The plan of the paper is as follows. To envisage a realistic situation
we will assume a slab of a homogeneous dielectric medium in vacuum
(thus we have two boundaries) and write down the corresponding plane
wave solutions of Maxwell's equations. These solutions will enable us
to quantize the electromagnetic field in the presence of the medium.
In the discussion of the probability for the emission of a photon
when the particle crosses the slab we will specialize to neutral
particles by choosing the appropriate form factors of the electromagnetic
current. The main result of this paper will be obtained by deriving an
exact but simple formula for the above probability in the ultrarelativistic
limit of zero particle masses and the case of the incident particle
momentum being orthogonal to the surface of the dielectric layer.
Finally we will evaluate this formula at photon energies above
$\hbar \omega_p$ where $\omega_p$ is the plasma frequency. Finally
neutrinos will come into the game by using the solar neutrino flux in
the numerical example and taking into account the experimental restrictions on
$\mu^2 + d^2$.

As already mentioned we envisage a situation where there is a layer of a
medium with dielectric constant $\ve = n^2 \neq 1$ ($n$ is the refraction
index) occupying the space defined by $|z| < a/2$. Thus the surfaces
of the layer are orthogonal to the $z$--axis. For simplicity we assume
vacuum outside the layer and that the permeability of the medium is 1. For the
description we need the following wave vectors:
$$
\vec k = \omega \left( \ba{c} \sin \alpha \cos \phi \\
\sin \alpha \sin \phi \\ \pm \cos \alpha \ea \right) , \qquad
\vec k_m = n \omega \left( \ba{c} \sin \beta \cos \phi \\
\sin \beta \sin \phi \\ \pm \cos \beta \ea \right) ,
$$
\beq
\vec k_r = S \vec k, \qquad
\vec k_{mr} = S \vec k_m \qquad \mbox{with} \qquad
S = \mbox{diag }(1,1,-1).
\eeq
$\vec k$ denotes the wave vector of incident and transmitted waves
whereas $\vec k_m$ pertains to the refracted solution inside the medium.
In addition, to fulfill Maxwell's equations there are reflected waves
with $\vec k_r$ in that part of the space from where the wave is coming
and $\vec k_{mr}$ in the medium. The angles of incidence $\alpha$ and
of refraction $\beta$ are related by Snell's law
\beq
\sin \alpha = n \sin \beta
\eeq
and therefore $k_{x,y} = k_{mx,y}$. Since $\cos \alpha > 0$ the plus and
minus signs in $k_z$ refer to waves incident from the spaces
$z < - a/2$ and $z > a/2$, respectively.

It is easy to show that given a solution $\vec E$, $\vec B$ of
Maxwell's equation for the above situation then the fields
\beq
\vec E'(\vec x,t) = S \vec E(S \vec x,t), \qquad
\vec B'(\vec x,t) = - S \vec B(S \vec x,t) \label{EB}
\eeq
form another solution because the medium and its position in space is
symmetric with respect to $z \ra - z$. Consequently, plane waves with
$k_z < 0$ can be obtained from solutions with $k_z > 0$ by applying
(\ref{EB}). Therefore we will confine ourselves to $k_z > 0$ for the
time being.

As a polarization basis for the electric field of the incident wave we will
choose
\beq
\vec e_I = \left( \ba{c} - \sin \phi \\ \cos \phi \\ 0 \ea \right)
\qquad \mbox{and} \qquad
\vec e_{II} = \left( \ba{c} \cos \alpha \cos \phi \\
\cos \alpha \sin \phi \\ - \sin \alpha \ea \right) =
- \frac{1}{\omega} \vec k \times \vec e_I,
\eeq
and we will search for solutions $\vec E$, $\vec B$ with time dependence
given by $e^{-i \omega t}$. In such a situation the magnetic field is
simply obtained by
\beq
\vec B = \frac{1}{i \omega} \mbox{ curl } \vec E . \label{B}
\eeq
It will turn out later that in the calculation of the transition
radiation it is sufficient to know the electric field. Thus we
content ourselves with describing $\vec E_j$ $(j = I,II)$ in detail:
\beqa
\label{Fresnel}
\vec E_I(\vec k,x) &=& e^{-i \omega t} \vec e_I \cdot \left\{
\ba{ll} e^{i \vec k \cdot \vec x} + a^I_r e^{i \vec k_r \cdot \vec x},
& z < - a/2 \\[5pt]
a^I_m e^{i \vec k_m \cdot \vec x} + a^I_{mr} e^{i \vec k_{mr} \cdot
\vec x}, & -a/2 < z < a/2 \\[5pt]
a^I_t e^{i \vec k \vec x}, & z > a/2 \ea \right. \no \\[5pt]
\vec E_{II}(\vec k,x) &=& e^{- i \omega t} \cdot \left\{
\ba{ll} \vec e_{II} e^{i \vec k \cdot \vec x} + \vec e_{IIr} a^{II}_r
e^{i \vec k_r \cdot \vec x}, & z < - a/2 \\[5pt]
\vec e_{IIm} a^{II}_m e^{i \vec k_m \cdot \vec x} +
\vec e_{IImr} a^{II}_{mr} e^{i \vec k_{mr} \cdot \vec x}, &
- a/2 < z < a/2 \\[5pt]
\vec e_{II} a^{II}_t e^{i \vec k \cdot \vec x}, & z > a/2 . \ea \right.
\eeqa
The additional polarization vectors in $\vec E_{II}$ are defined by
\beqa
\vec e_{IIm} &=& \left( \ba{c} \cos \beta \cos \phi \\
\cos \beta \sin \phi \\ - \sin \beta \ea \right) =
- \frac{1}{n \omega} \vec k_m \times \vec e_I, \no \\
\vec e_{IIr} &=& - \frac{1}{\omega} \vec k_r \times \vec e_I =
- S \vec e_{II}, \\
\vec e_{IImr} &=& - \frac{1}{n \omega} \vec k_{mr} \times \vec e_I =
- S \vec e_{II m}. \no
\eeqa
Continuity of $\vec E_\parallel$ and $\bar \ve \vec E_\perp$ at the
surfaces $z = \pm a/2$ determines the coefficients in (\ref{Fresnel})
where
\beq
\bar \ve (z) = \left\{ \ba{cc} \ve, & |z| < a/2 \\[5pt]
1, & |z| > a/2 \ea \right.
\eeq
describes the variation of the dielectricity constant in space. With
the definitions
\beq
K = e^{i|k_z|a/2}, \qquad
K_m = e^{i|k_{mz}|a/2}, \qquad
\rho = k_{mz}/k_z = n^2 \wt \rho ,
\eeq
and
\beq
\N = (1 + \rho)^2 K^*_m{}^2 - (1 - \rho)^2 K^2_m, \qquad
\wt \N = (1 + \wt \rho)^2 K^*_m{}^2 - (1 - \wt \rho)^2 K^2_m ,
\eeq
one can write down the list of coefficients:
\beq
\ba{lcllcl}
a^I_r &=& (1 - \rho^2) K^*{}^2 (K^*_m{}^2 - K^2_m)/\N, &
a^I_t &=& 4 \rho K^*{}^2/\N, \\[5pt]
a^I_m &=& 2(1 + \rho) K^* K^*_m/\N, &
a^I_{mr} &=& - 2(1 - \rho) K^* K_m/\N ; \\[5pt]
a^{II}_r &=& (1 - \wt \rho^2) K^*{}^2 (K^*_m{}^2 - K^2_m)/\wt \N, &
a^{II}_t &=& 4 \wt \rho K^*{}^2/\wt \N, \\[5pt]
a^{II}_m &=& 2(1 + \wt \rho) K^* K^*_m/n \wt \N, &
a^{II}_{mr} &=& - 2(1 - \wt \rho) K^* K_m/n \wt \N .
\ea
\eeq
It can be checked that (\ref{Fresnel}) and (\ref{B}) are normalized
like the vacuum solutions to
\beq
\frac{1}{2} \int d^3x (\bar \ve(z) \vec E_j(\vec k,x)^* \cdot
\vec E_{j'}(\vec k',x) + \vec B_j(\vec k,x)^* \cdot
\vec B_{j'}(\vec k',x)) = (2 \pi)^3 \delta_{jj'} \delta(\vec k -
\vec k')
\eeq
and that the corresponding integrals without complex conjugation
give zero. This allows for the straightforward quantization of
\beq
\vec E(x) = \sum_{j=I,II} \int \frac{d^3k}{(2\pi)^{3/2}}
\sqrt{\frac{\omega}{2}} (a_j(\vec k) \vec E_j(\vec k,x) +
a^\dg_j(\vec k) \vec E_j(\vec k,x)^*)
\eeq
by
\beq
[a_j(\vec k),a^\dg_{j'}(\vec k')] = \delta_{jj'} \delta(\vec k - \vec k'),
\qquad [a_j(\vec k), a_{j'}(\vec k')] = 0
\eeq
and consequently
\beq
H_\gamma = \frac{1}{2} \int d^3 x: (\bar \ve(z) \vec E{}^2(x) +
\vec B{}^2(x)) : = \sum_{j=I,II} \int d^3 k \, \omega a^\dg_j(\vec k)
a_j(\vec k).
\eeq

Now we turn to the calculation of the probability of
\beq
\nu_i(p_i,s_i) \ra \nu_f(p_f,s_f) + \gamma(k,j) \qquad (j = I,II)
\label{rad}
\eeq
in the presence of the dielectric layer. Here $\nu_i$, $\nu_f$ stand
for any neutral fermion with magnetic and electric (transition) moments
$\mu$, $d$, respectively. The relation \cite{be}
\beq
i(E_i - E_f) \int d^4 x A_\mu(x) \langle p_f,s_f|J^\mu(x) | p_i,s_i\rangle
= \int d^4x \vec E(x) \cdot \langle p_f,s_f | \vec J(x) | p_i,s_i\rangle,
\eeq
where $J^\mu$ is the electromagnetic current and $A_\mu$ the vector
potential, greatly simplifies the calculation. In our case we have
\beq
\langle p_f,s_f | J^\mu(x) | p_i,s_i\rangle = \frac{1}{(2\pi)^3}
\left( \frac{m_f m_i}{E_f E_i} \right)^{1/2} e^{-iq \cdot x}
\bar u_f (- i \mu + d \gamma_5) \sigma^{\mu\nu} q_\nu u_i
\label{formfac}
\eeq
with $q = p_i - p_f$ and obvious abbreviations $u_i$, $u_f$. Assuming an
initially polarized fermion but summing over the final polarization $s_f$
we obtain
\beqa
\label{diff}
d^6 W &=& \sum_{s_f} \sum_{j=I,II}
|\bar u_f(\mu + i \gamma_5 d) \sigma_{\mu\nu} {\cal E}_j^{\mu*} q^\nu u_i/
\omega|^2 \cdot
 (2\pi)^3 \delta(p_i^1 - p^1_f - k^1) \cdot \no \\
&& \cdot \delta(p_i^2 - p_f^2 - k^2)
\delta (E_i - E_f - \omega) \cdot \frac{m_i}{p^3_i} \cdot
\frac{m_f}{(2\pi)^3 E_f} d^3 p_f \cdot \omega^2
\frac{d^3 k}{(2\pi)^3 2 \omega}
\eeqa
with
\beq
({\cal E}^\mu_j) = \left( \ba{c} 0 \\ \vec {\cal E}_j \ea \right) \quad
\mbox{and} \quad
\vec {\cal E}_j(\vec k,q_z) = \left. \int \! dz \, e^{-iq_zz}
\vec E_j(\vec k,x)
\right|_{x^0 = x^1 = x^2 = 0}
\eeq
for the probability of the process (\ref{rad}). The $\delta$--functions
reflect the symmetries of the problems leading to energy conservation and
momentum conservation in the $xy$--plane which allows to express $p_f$ and
$q$ as functions of $p_i$ and $k$:
\beq
p_f = \left( \ba{c} E_i - \omega \\ p^1_i - k^1 \\ p^2_i - k^2 \\ \eta P
\ea \right), \quad
q = \left( \ba{c} \omega \\ k^1 \\ k^2 \\ p^3_i - \eta P \ea \right)
\quad \mbox{with} \quad
P = [(E_i - \omega)^2 - m^2_f - (p^1_i - k^1)^2 - (p^2_i - k^2)^2]^{1/2}.
\eeq
$\eta = \pm 1$ corresponds to forward and backward scattering of the
fermion, respectively.

In the computation of $\vec {\cal E}_j$ $(j = I,II)$ it turns out that
$a_r^j$ and $a_t^j$ can suitably be expressed by $a^j_m$ and $a^j_{mr}$
such that these latter coefficients appear only in the expressions
\beq
S^j_- \equiv a^j_m \frac{\sin \frac{a}{2}(k_{mz} - q_z)}{k_{mz} - q_z},
\qquad
S^j_+ \equiv a^j_{mr} \frac{\sin \frac{a}{2} (k_{mz} + q_z)}
{k_{mz} + q_z} .
\eeq
Finally one arrives at
\beqa
\vec {\cal E}_I(\vec k,q_z) &=& 2(S^I_- + S^I_+)(1 - n^2) \frac{\omega^2}
{k^2_z - q^2_z} \vec e_I, \no \\
\vec {\cal E}_{II}(\vec k,q_z) &=& 2 \sigma(\theta) \left(\frac{1}{n} - n
\right) \frac{1}{k^2_z - q^2_z} \cdot \no \\
&& \cdot \{ [ S_-^{II} (q_z \omega \sin^2 \theta + k_{mz} k^2_z/\omega)
+ S_+^{II} (q_z \omega \sin^2 \theta - k_{mz} k^2_z/\omega)] \vec e_\phi +
\no \\
&& \mbox{} + [S_-^{II} (q^2_z - \omega^2 - q_z k_{mz}) +
S_+^{II} (q^2_z - \omega^2 + q_z k_{mz})] \sin \theta \vec e_z\}.
\eeqa
The two unit vectors in $\vec {\cal E}_{II}$ are defined as
\beq
\vec e_\phi = \left( \ba{c} \cos \phi \\ \sin \phi \\ 0 \ea \right),
\qquad
\vec e_z = \left( \ba{c} 0 \\ 0 \\ 1 \ea \right).
\eeq
Instead of the angle of incidence $\alpha$ we now use
$\theta = < \! \! \! ) (\vec k,\vec e_z)$ for convenience to subsume forward
$(k_z > 0)$ and backward $(k_z < 0)$ moving incident photons. Thus
\beq
k_z = \omega \cos \theta, \qquad
k_{mz} = \sigma(\theta) \omega \sqrt{n^2 - \sin^2\theta} \quad
\mbox{with} \quad \sigma(\theta) = \frac{\cos \theta}{|\cos \theta|}
\eeq
and $0 \leq \theta \leq \pi$ ($\theta \neq \pi/2$). Note that the
coefficients $a^j_m$, $a^j_{mr}$ do not change under
$\theta \ra \pi - \theta$.

Up to now no approximations have been made. In the following, however,
we will evaluate (\ref{diff}) in the ultrarelativistic limit. To this
end we take advantage of the Gordon decomposition and obtain
\beq
4m_i m_f \sum_{s_f} | \bar u_f (\mu + i \gamma_5 d) \sigma_{\mu\nu}
{\cal E}^{\mu*} q^\nu u_i |^2 \ra (\mu^2 + d^2)(q^2_z - k^2_z)|
(\vec p_f + \vec p_i) \cdot \vec {\cal E}|^2
\eeq
for $m_i,m_f \ra 0$. Let us stress three consequences of this limit:
\begin{enumerate}
\item[i)] The form of the dipole moment interaction (\ref{formfac})
requires a spin flip of the fermion. Consequently, the classical
treatment of transition radiation is not applicable in this
case.\footnote{In the classical approach, one calculates the radiation
emitted by a magnetic (or electric) dipole moving with constant velocity
and pointing towards a fixed direction in space. Therefore, in the
above limit, the classical calculation gives a zero result \cite{sa}.}
\item[ii)] The transition radiation is independent of the polarization
$s_i$ of the incident fermion.
\item[iii)] For $\vec p_i = E_i \vec e_z$ the polarization $I$ of the
photon does not contribute and the probability (\ref{diff}) is
independent of the angle $\phi$.
\end{enumerate}
Adopting iii) in addition to $m_i,m_f \ra 0$ we thus obtain for the
probability (\ref{diff})
\beqa
\label{res}
d^2W &=& \sum_{\eta = \pm 1} \frac{\sin^3\theta d\theta \omega^3 d\omega}
{2 \pi^2} \frac{E}{P} (\mu^2 + d^2)\left( n - \frac{1}{n}\right)^2 \cdot
\no \\
&& \cdot \frac{1}{q^2_z - k^2_z} |S_-^{II}(q_z - \omega - k_{mz}) +
S_+^{II} (q_z - \omega + k_{mz})|^2
\eeqa
with
\beq
E \equiv E_i, \qquad
q_z = E - \eta P, \qquad
P = (E^2 - 2E\omega + \omega^2 \cos^2\theta)^{1/2}.
\eeq
This is the main result of the present work.

We will now apply (\ref{res}) to the case where the photon energy
$\omega$ is much larger than the plasma frequency $\omega_p$ of the
medium\footnote{Polypropylene with $\omega_p = 20$~eV may serve as a
typical example.}. In this energy range, the dielectric constant is
given by $\ve = n^2 = 1 - \omega^2_p/\omega^2$. The differential
production rate is dominated by the kinematical region where $k_z
\simeq q_z$, corresponding to small values of the angle $\theta$.
This is a consequence of $a \omega \gg 1$ for realistic situations
(for example $a \omega \simeq 10^6$ for $a=10^{-3}$~cm and
$\omega=20$~keV). For the same reason also $q_z \simeq \omega +
\omega^2 \theta^2/2(E-\omega)$ holds for all values of
$\omega$ except for a small and negligible range close to $E$.
Furthermore, contributions from backscattering of the photon or the fermion
in (\ref{res}) are also completely negligible and $|a^{II}_m|
\simeq 1$, $a^{II}_{mr} \simeq 0$. Thus to a very
good approximation, the final result is then given by
\beq
d^2W \simeq \frac{2(\mu^2 + d^2)}{\pi^2} d\omega \; \omega \; d \theta \;
\theta \left( \frac{\omega_p}{\omega}\right)^4
\left\{ \frac{\sin \frac{a \omega}{4} [( \frac{\omega_p}{\omega})^2
+ \theta^2 \frac{E}{E - \omega}]}
{(\frac{\omega_p}{\omega})^2 + \theta^2 \frac{E}{E - \omega}}\right\}^2.
\eeq
This leads to the photon spectrum
\beq
dW \simeq \frac{(\mu^2 + d^2)a}{4 \pi^2} d\omega \; \omega^2
\frac{E - \omega}{E} \left( \frac{\omega_p}{\omega}\right)^4
F \left( \frac{a \omega^2_p}{4 \omega}\right),
\eeq
where
\beq
F(u) = \int_u^\infty dx \; \frac{\sin^2 x}{x^2} .
\eeq

We consider the situation where $a \ll 4 \omega/\omega^2_p$. (For
detectors of comparable size, the opposite case of large $a$ leads to a
much smaller event rate.)
In this case $F(u)$ can be approximated by $F(0) = \pi/2$, and the
total emission probability in the photon energy range
$\omega_{\rm min} \leq \omega \leq E$ is given by
\beq
W \simeq \frac{(\mu^2 + d^2)a \omega^4_p}{8\pi} \left[
\frac{1}{\omega_{\rm min}} - \frac{1}{E} \left( 1 + \ln
\frac{E}{\omega_{\rm min}} \right) \right]
\simeq \frac{(\mu^2 + d^2) a \omega^4_p}{8 \pi \omega_{\rm min}}.
\label{W1}
\eeq
For our numerical example we take $\omega_{\rm min} = 20$~keV and
$\omega_p = 20$~eV. With these values, formula (\ref{W1}) is valid for
$a \ll 4 \cdot 10^{-3}$~cm. The total number of events for a flux of
incoming particles $I$, a detector cross section $A$ and $N$ foils
during a time interval $T$ is given by
\beq
\#(\mbox{events}) = WIANT .
\eeq
As we are interested in the transition radiation produced by solar
neutrinos, we normalize our result to $I = 6
\cdot 10^{10}$~cm$^{-2}$~s$^{-1}$, the solar neutrino flux expected
from the standard solar model \cite{ba}. The quantity $\mu^2 + d^2$
in (\ref{W1}) has to be understood as some sort of ``effective
magnetic moment'' \cite{gri},
\beq
\mu^2_{\rm eff} = \sum_{i,j} (|\mu_{ji}|^2 + |d_{ji}|^2) p_i,
\eeq
where $p_i$ is the probability to find a neutrino with flavour $i$ in
the solar neutrino flux.
We normalize $\mu_{\rm eff}$ to $10^{-9} \mu_B$. The remaining
quantities are normalized to $a = 4 \cdot 10^{-4}$~cm,
$A = 100$~m$^2$, $N = 10^6$ and $T = 1\mbox{ yr} \simeq 3 \cdot 10^7$~s.
Then we obtain
\beq
\#(\mbox{events}) \simeq \left( \frac{a}{4 \cdot 10^{-4}\mbox{ cm}}
\right) \left( \frac{\mu_{\rm eff}}{10^{-9} \mu_B}\right)
\left( \frac{A}{100\mbox{ m}^2}\right) \left( \frac{T}{1\mbox{ yr}}
\right) \left( \frac{N}{10^6}\right).
\label{ev}
\eeq

Unfortunately, this event rate is extremely small leaving little hope
to detect a neutrino MM by the mechanism of transition radiation.
However, one remaining possibility would be the case of a large $p_\tau$
and a $\mu_{\tau\tau}$ near its present upper bound.

In concluding this paper we want to mention a few points. In the
numerical calculation for a stack of $N$ foils we have simply
multiplied the probability for one foil by $N$. This is certainly not
correct if the distance between the foils becomes comparable to their
thickness and thus interference effects become important. However, if
the situation considered here is similar to the transition radiation of
electrons in this respect then we cannot expect a substantial increase
of the radiation yield \cite{du} and equ. (\ref{ev}) still holds as an
order of magnitude estimate. Furthermore, we have not taken into
account total reflexion but for $\omega \gg \omega_p$ its effect will
be negligible. On the other hand, for large $n-1$ one could have such
geometries of the dielectric medium where total reflexion is important.
Finally, we want to stress once more that the methods used in our
calculation are quite general and could thus be carried over to cases
where one is not satisfied with the usual approximations made in the
field of transition radiation.

\end{document}